\documentclass[10pt]{article}
\usepackage{graphicx}
\usepackage{subfigure}
\begin{document}
\pagenumbering{arabic}
\def\ao{\hat{a}}
\def\aot{\hat{a}^2}
\def\co{\hat{a^\dagger}}
\def\no{\co\ao}
\def\cs{\vert\alpha\rangle}
\def\ncs{\vert-\alpha\rangle}
\def\ecs{\vert\alpha +\rangle}
\def\ocs{\vert\alpha -\rangle}
\def\bcs{\vert\beta\rangle}
\def\bncs{\vert(-1)^{1\over k}\beta\rangle}
\def\becs{\vert\beta +\rangle}
\def\bocs{\vert\beta -\rangle}
\def\pcs{\vert +\rangle}
\def\mcs{\vert -\rangle}
\def\sech{\hbox{sech}}
\bibliographystyle{unsrt}
\title{Entanglement in bipartite generalized coherent states}
\author{S. Sivakumar\\Materials Science Division\\ Indira Gandhi Centre for Atomic Research\\
Kalpakkam 603 102 INDIA\\
{Email: siva@igcar.gov.in}}
\maketitle
\begin{abstract}

	Entanglement in a class of bipartite generalized coherent states is discussed.  It is shown that a positive parameter can be associated with the bipartite generalized coherent states so that the states with equal value for the parameter are of equal entanglement. It is shown  that the maximum possible entanglement of 1 bit is attained if the positive parameter equals $\sqrt{2}$. The result that the entanglement is one bit when the relative phase between the composing states is $\pi$ in bipartite coherent states is shown to be true for the class of bipartite  generalized coherent states considered.  
\end{abstract}
PACS : 42.50.Dv;03.65.Ca;03.67.Mn\\
Keywords: Entanglement, generalized coherent states, bipartite entanglement
\newpage
\section{Introduction}

Entanglement in a quantum system is possible  when the  system has more than one degree of freedom.  A bipartite system has two degrees of freedom.  The two degrees of freedom may correspond to two physically different systems, say, the spins of two spin-half particles or entanglement may exist between two degrees of freedom of a same particle, for instance, a particle moving in two dimensions.  Associated with every degree of freedom is a Hilbert space. The state of the composite system is described by the vectors in the direct product space of the Hilbert spaces corresponding to the various degrees of freedom.  The importance of entanglement stems from the fact that if two particles are in an entangled state, then it is not possible  to describe the state of the participating systems separately.  An important example is a system of two spin-half particles, the best known example of a two-level system. Quantum computers are envisaged to be built with 2-level systems where entanglement plays a crucial role in making these gadgets more efficient than the classical computing machines\cite{nielsen}. A possible set of basis vectors, referred in literature as Bell basis,  to describe a system of two spin-half particles is 
\begin{eqnarray}
\vert e1\rangle&=&{1\over\sqrt{2}}\left[\vert\uparrow,\uparrow\rangle+\vert\downarrow,\downarrow\rangle\right],\\
\vert e2\rangle&=&{i\over\sqrt{2}}\left[\vert\uparrow,\uparrow\rangle-\vert\downarrow,\downarrow\rangle\right],\\
\vert e3\rangle&=&{i\over\sqrt{2}}\left[\vert\uparrow,\downarrow\rangle+\vert\downarrow,\uparrow\rangle\right],\\
\vert e4\rangle&=&{1\over\sqrt{2}}\left[\vert\uparrow,\downarrow\rangle-\vert\downarrow,\uparrow\rangle\right].
\end{eqnarray}
These states possess a host of important features.    In the context of the present work, the following result of Bennet {\it et al}\cite{bennet-prl-1996}, which was later extended to bipartite mixed states\cite{bennet-pra-1996,wootters-prl-1997},  for pure states is crucial.  The result quantifies the amount of entanglement in a bipartite system.  
If a bipartite pure state is expressed in the Bell basis as $\sum_{j=1}^{4}c_j\vert ej\rangle$, the {\it concurrence} $c$  between the subsystems is  $\vert\sum_{j=1}^{4}c_j^2\vert\le 1$.  Defining $x=.5+.5\sqrt{1-c^2}$, the entanglement $E$ associated with the bipartite system is 
\begin{equation}\label{entang}
E=-\left[ x\log_2 x+(1-x)\log_2(1-x)\right]
\end{equation}
Entanglement decreases from one bit to zero monotonically as $x$ increases from 0.5 to 1.   For a separable bipartite state the value of $E$ is zero. All other values imply that there is entanglement between the subsystems  of the bipartite system.  It is to be noted that the maximum value of entanglement is unity, corresponding to $x=0.5$, equivalently $c=1$.  All the states of the Bell basis have one bit of entanglement.  Though originally derived for systems whose Hilbert spaces are of dimension two, it is possible to extend this idea to some special states in infinite dimensional cases. For composite systems whose Hilbert space dimensions exceed two, a necessary condition for separability, {\it i.e.}, zero entanglement, is the positivity of the partial transpose of the density matrix for the system\cite{peres}.  This has been shown to be sufficient if the dimension of one subsystem is two and the other is two or three\cite{horo1,horo2}.  In the present  work, the states considered belong to infinite dimensional Hilbert spaces; however, the choice of parameters in the bipartite superpositions allows for analysis in a basis that is two-dimensional for each subsystem. 

A coherent state $\vert\alpha\rangle$ is the eigenstate of the harmonic oscillator annihilation operator $\ao$. It is defined  for any $\alpha$ in the complex plane,  by the following expansion in the number state basis:
\begin{equation}
\vert\alpha\rangle=\exp(-{\vert\alpha\vert^2\over 2})\sum_{n=0}^\infty\frac{\alpha^n}{\sqrt{n!}}\vert n\rangle.
\end{equation}
These states belong to the infinite dimensional Hilbert space spanned by the number states of the harmonic oscillator. 
The coherent states correspond to Gaussian states in position representation and the dynamics of  any observable is the same as that of the corresponding classical variable.  By superposing two coherent states, namely, $\cs$ and $\ncs$, two orthogonal states
\begin{eqnarray}
\vert \alpha +\rangle &=& n_+[\cs+\ncs],\label{ecs}\\
\vert \alpha -\rangle &=& n_-[\cs -\ncs],\label{ocs}
\end{eqnarray}
are constructed.   These orthogonal states are normalizable for any value of $\alpha$. The constants $n_\pm$ are to ensure that the states are normalized.  The states $\cs$ and $\ncs$ can be expressed as linear combinations of these two orthogonal states.  It is to be noted that the orthogonal states constructed out of coherent states of amplitudes $\alpha$ and $-\alpha$ are of no use to express a coherent state of different amplitude.  The basis given in Eqs. \ref{ecs} and \ref{ocs} is specific to a particular amplitude.   

Two-mode mode coherent states of the form $\cs\cs+\exp(i\phi)\ncs\ncs$ are amenable to analysis using Bell-type coherent states constructed out of two-mode orthogonal states defined below:
\begin{eqnarray}
\vert E1\rangle&=&{1\over\sqrt{2}}\left[\vert +,+\rangle+\vert -,-\rangle\right],\\
\vert E2\rangle&=&{i\over\sqrt{2}}\left[\vert +,+\rangle-\vert -,-\rangle\right],\\
\vert E3\rangle&=&{i\over\sqrt{2}}\left[\vert +,-\rangle+\vert -,+\rangle\right],\\
\vert E4\rangle&=&{1\over\sqrt{2}}\left[\vert +,-\rangle-\vert -,+\rangle\right].
\end{eqnarray}
The symbol $\vert +,+\rangle$ stands for the two-mode state $\ecs\ecs$ and similar expressions are associated with the other two-mode states. These basis states have the same structure as the Bell basis for the bipartite spin-1/2 system.  Hence, the analysis of both the systems proceed along the same lines.  Apart from an overall normalization factor, the  two-mode coherent state is expressed in this basis as  
\begin{equation}\label{2mcs}
\cs\cs+\exp(i\phi)\ncs\ncs = \frac{1+\exp(i\phi)}{\sqrt{2}}\left[\vert E1\rangle +i\exp(-2\vert\alpha\vert^2)\vert E2\rangle\right] +i\sqrt{1-\exp(-4\vert\alpha\vert^2)}(\exp(i\phi)-1) \vert E3\rangle.
\end{equation}
Here $\phi$ is the relative phase between the states. The previous equation defines the type of two-mode coherent states which can be written down using the Bell basis for the coherent states. 
The entanglement associated with this two-mode coherent state is given by the expression defined in Eq. \ref{entang} and the coefficients of the basis states $\vert Ej\rangle\}_{j=1}^4$ in Eq. \ref{2mcs} are to be used for $c_j$ in computing $x$.  Interestingly, when $\phi=\pi$, the entanglement is exactly one bit and independent of $\alpha$\cite{sasaki-2001}.  In the present  work, the result is shown to hold for a class of two-mode generalized coherent states.  Further,  it is established that one bit of entanglement for all values of $\phi$ is possible by a proper choice of the amplitude defining a generalized  coherent state.  The results are discussed in the context of coherent states (CS), squeezed vacuum (SV)\cite{knight}, even coherent states (ECS), odd coherent states (OCS)\cite{dodonov} and logarithmic states (LS)\cite{mvs}.

\section{Bell basis for generalized coherent states}

The coherent states $\cs$  are associated with the harmonic oscillators.  In the  interaction of electromagnetic waves with nonlinear media, a variety of other states of the field are generated.  Often, such states are either eigenstates of a suitable annihilation operator or produced by the action of an unitary operator on the ground state.  Many such  states known in the literature are of the form
\begin{equation}
\vert\beta\rangle=\sum_{n=0}^{\infty}C_{kn}\beta^{kn}\vert nk + m\rangle\hspace{1cm}\beta\in C.
\end{equation}
The parameter $\beta$ is the amplitude of the generalized coherent state.  
The coefficients $C_{kn}$ are functions of $n$ and $\vert\beta\vert$. The coefficients satisfy  $\sum_{n=0}^{\infty}\vert C_{kn}\beta^{kn}\vert^2=1$, the usual normalization condition. The parameters $k$ and $m$ are positive integers.   The parameters for some of the known coherent states are given in Table 1.  In the case of logarithmic states, the amplitude $q$ is real and its magnitude is less than unity.  The parameter $\gamma$ is complex and unrestricted.
\begin{table}[ht]
\caption{$k$, $m$ and $c_n$ for various coherent states}
	\begin{center}
		\begin{tabular}{|l||c|c|l|}
\hline
State  & $k$&  $m$ & $c_{kn}$\\
\hline
CS & 1 & 0 & $\frac{\exp(-{\vert\beta\vert^2\over 2})}{\sqrt{n!}}$\\
\hline
SV & 2 & 0&$\frac{\sqrt{(kn)!}}{n!\sqrt{\cosh(\vert\beta\vert^2)}}\left[\frac{\tanh(\vert\beta\vert^2)}{2\vert\beta\vert^2}\right]^{n}$\\
\hline
ECS & 2 & 0 &${1\over\sqrt{\cosh(\vert\beta\vert^2)(kn)!}}$\\
\hline
OCS & 2 & 1&${\vert\beta\vert\over\sqrt{\sinh(\vert\beta\vert^2)(kn+1)!}}$\\
\hline
LS & 1 & 0 & $\gamma\over\sqrt{\vert\gamma\vert^2-\log(1-\vert q\vert^2)}$ for $n=0$ \\
& & & $\frac{1}{n\sqrt{\vert\gamma\vert^2-\log(1-\vert q\vert^2)}}$ for $n > 0$\\
\hline
		\end{tabular}
	\end{center}
\label{states}
\end{table}

Similar to $\ecs$ and $\ocs$ for the coherent states $\cs$, the following orthogonal states are defined for the generalized coherent states $\vert\beta\rangle$:
\begin{eqnarray}
\vert\beta +\rangle&=&N_+[\vert\beta\rangle + \vert (-1)^{1\over k}\beta\rangle]=A\sum_{n=0}^{\infty}C_{2nk}\beta^{2n}\vert 2nk+m\rangle,\\
\vert\beta -\rangle&=&N_-[\vert\beta\rangle - \vert (-1)^{1\over k}\beta\rangle]=B\sum_{n=0}^{\infty}C_{(2n+1)k}\beta^{2n+1}\vert (2n+1)k+m\rangle.
\end{eqnarray}
It is to be noted that the states involved in the definition of the orthogonal states correspond to amplitudes $\beta$ and $(-1)^{1\over k}\beta$.   For the CS,  $k=1$ and hence the $\bcs$ and $\bncs$ correspond to $\cs$ and $\ncs$.  
The constants $N_+$ and $N_-$ are normalization coefficients. The parameters $A$ and $B$  are positive and given by the expressions,
\begin{eqnarray}
A &=&\left[\sum_{n=0}^{\infty}\vert C_{2nk}\beta^{2nk}\vert^2\right]^{-{1\over 2}},\\
B&=&\left[\sum_{n=0}^{\infty}\vert C_{(2n+1)k}\beta^{(2n+1)k}\vert^2\right]^{-{1\over 2}}.
\end{eqnarray}
Since the generalized coherent states are normalized the parameters satisfy $A^{-2}+B^{-2}=1$.   The generalized coherent state $\vert\beta\rangle$ can be written a linear combination of $\bcs$ and $\bncs$.  AS in the case of CS, the basis provided by $\becs$ and $\bocs$ is suitable to expand a generalized coherent state whose  amplitude is either $\beta$ or $(-1)^{1/k}\beta$.  This basis of dimension two facilitates a description of entanglement in two-mode generalized coherent states that parallels the description for a system of two spin-1/2 particles. 

\section{Entanglement in two-mode generalized coherent states}
	
Any two-mode generalized coherent state of the form 
\begin{equation}\label{2modecs}
\vert\beta,(-1)^{1\over k}\beta\rangle=N\left[\bcs\bcs+\exp(i\phi)\bncs\bncs\right],
\end{equation} 
is expressible in terms of the two-mode extensions of $\vert\beta+\rangle$ and $\vert\beta-\rangle$.  The normalization constant $N$ is $1/\sqrt{2[1+X^2\cos\phi]}$, where $X^2=2A^{-2}-1$.  The Bell basis states to express two-mode generalized coherent states  are
\begin{eqnarray}
\vert G1\rangle&=&{1\over\sqrt{2}}\left[\becs\becs+\bocs\bocs\right],\\
\vert G2\rangle&=&{i\over\sqrt{2}}\left[\becs\becs-\bocs\bocs\right],\\
\vert G3\rangle&=&{i\over\sqrt{2}}\left[\becs\bocs+\bocs\becs\right],\\
\vert G4\rangle&=&{1\over\sqrt{2}}\left[\becs\bocs-\bocs\becs\right].
\end{eqnarray}

If a state has expansion coefficients $\{\alpha_j\}_{j=1}^{4}$ in the Bell basis $\{\vert Gj\rangle\}_{j=1}^{4}$   and  $\{a_j\}_{j=1}^{4}$ in the basis $\{\becs\becs,\bocs\bocs,\becs\bocs,\bocs\becs\}$, then 
\begin{eqnarray}
\alpha_1&=&{a_1+a_2\over\sqrt{2}},\\
\alpha_2&=&i{a_1-a_2\over\sqrt{2}},\\
\alpha_3&=&i{a_3+a_4\over\sqrt{2}},\\
\alpha_4&=&{a_3-a_4\over\sqrt{2}}.
\end{eqnarray}
In terms of the coefficients $a_j$, the concurrence $c$, given by $\vert\sum_{j=1}^{4}\alpha_j^2\vert$,  is expressed as a determinant,
\begin{equation}\label{deter}
c=2{\left|\left|
\begin{array}{cc}
        a_1 & a_3\\
        a_4 & a_2  \\
     \end{array}\right|\right|.}
\end{equation}
The symbol $\left|\left|\cdot\cdot\right|\right|$ stands for the absolute value of the determinant.
When the relative phase $\phi$ is zero or $\pi$, the coefficients ${a_j}$ are real for the class of states $\vert\beta,(-1)^{1\over k}\beta\rangle$. In that case the concurrence has the interpretation as the  area of a parallelogram with vertices at the origin, $(a_1,a_3)$ and $(a_2,a_4)$.

The two-mode generalized coherent state expressed in the later basis is
\begin{eqnarray}
\vert\beta,(-1)^{1\over k}\beta\rangle &=& N[1+\exp(i\phi)]\left[\frac{1}{A^2}\becs\becs+\frac{1}{B^2}\bocs\bocs\right]\nonumber\\& &+N\frac{1-\exp(i\phi)}{AB}\left[\becs\bocs+\bocs\becs\right],
\end{eqnarray}
and the coefficients $a_j$ are easily obtained from the expression. 
The concurrence  between the two modes, computed using Eq. \ref{deter}, is
\begin{equation}
c = {1-X^2\over 1+X^2\cos\phi}.
\end{equation}
The above equation is the important result in this work.  All two-mode coherent states of the type defined by $\vert\beta,(-1)^{1\over k}\beta\rangle$ having equal value of $A$ posses equal entanglement.  Since the entanglement is a monotonic function of concurrence, it is enough to study the behaviour of the later.  In Fig. 1  the concurrence for the states defined in Eq. \ref{2modecs} is shown as a function of the parameter $A$ for different values of $\phi$.  The fact that the entanglement is one bit for any $\beta$ when $\phi=\pi$ is obvious from the expression for concurrence; $c$ becomes unity when $\phi=\pi$ and hence the entanglement is one bit.  
\begin{figure}[htp]
\centering
\includegraphics[width=10cm, height=8cm]{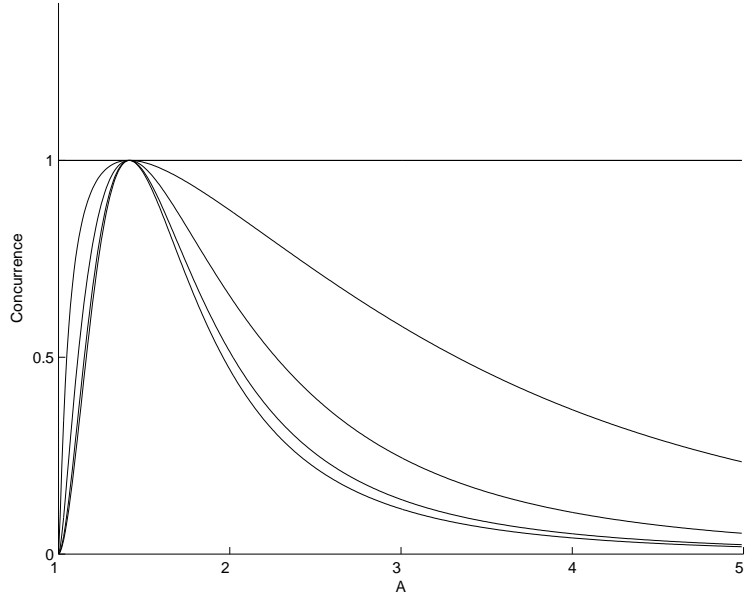}
\caption{Concurrence $c$ as a function of the parameter $A$.  The lowermost curve corresponds to $\phi=0$.  The curves above are for $\phi$ varied in steps of $0.25\pi$. The topmost curve corresponds to $\phi=\pi$ and in this case the concurrence is unity and independent of $A$ .}
\label{fig:Fig1}
\end{figure}

     The derivative of $x$ with respect to $X$ vanishes when $X=0$, which yields $A=\sqrt{2}$ for all values of $\phi.$  At this value of $A$, concurrence is unity and consequently the states $\vert\beta,(-1)^{1\over k}\beta\rangle$  have one bit of entanglement. In essence, the entanglement is one bit for all values of $\phi$ if $A=\sqrt{2}$.  This is complementary to the result that the entanglement is one bit for all value of $\beta$ if $\phi=\pi.$

   Different two-mode generalized coherent states with same value of $A$ have same concurrence and hence equal entanglement. However, for a given amplitude $\beta$, the value of $A$ differs for various states.  The $\vert\beta\vert$-dependence of the $A$ parameters for the different states are:
\begin{eqnarray}
A_{CS} &=& \sqrt{\frac{2}{1+\exp(-2\vert\beta\vert^2)}},\\
A_{SV} &=& \sqrt{\frac{2}{1+\sqrt{1-\tanh(2\vert\beta\vert^2)\tanh(\vert\beta\vert^2)}}},\\
A_{EC} &=& \sqrt{\frac{2\cosh(\vert\beta\vert^2)}{\cos(\vert\beta\vert^2)+\cosh(\vert\beta\vert^2)}},\\
A_{OC} &=& \sqrt{\frac{2\sinh(\vert\beta\vert^2)}{\sin(\vert\beta\vert^2)+\sinh(\vert\beta\vert^2)}},\\
A_{LS} &=&\sqrt\frac{\vert\gamma\vert^2-\log(1-\vert q\vert^2)}{\vert\gamma\vert^2-\frac{1}{2}\log(1-\vert q\vert^4)}.
\end{eqnarray}
The suffixes indicate the relevant states.  
The asymptotic value, obtained by letting $\vert\beta\vert\rightarrow\infty$ and $\vert q\vert\rightarrow 1$,  of $A$ in all the cases is $\sqrt{2}$. In Fig. 2 the dependence of $A$ on the magnitude of the complex amplitude $\beta$ is shown for the various  states given in Table. 1 except the logarithmic states.  
\begin{figure}[htp]
\centering
\includegraphics[width=10cm, height=8cm]{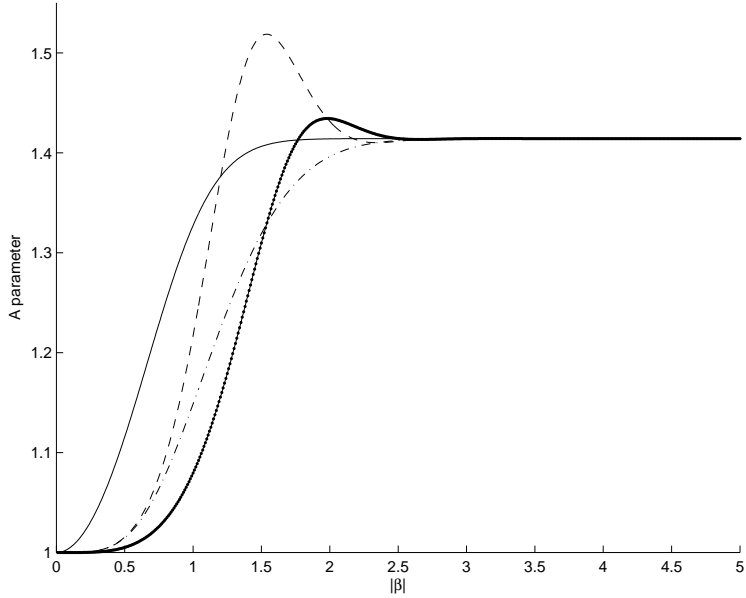}
\caption{Parameter $A$ as a function of $\vert\beta\vert$.  The curves shown correspond to  $\phi=0.5\pi$. The thin  continuous curve is for the two-mode coherent state, dashed curve for two-mode ECS, dashed-dotted curve for two-mode squeezed vacuum and the thick continuous line for two-mode OCS.}
\label{fig:Fig2}
\end{figure}
The concurrence in any of  the states given in Table. 1 becomes unity when  $A=\sqrt{2}$.   The entanglement associated with these states as a function of $\vert\beta\vert$ is given in Fig. 3.
\begin{figure}[htp]
\centering
\includegraphics[width=10cm, height=8cm]{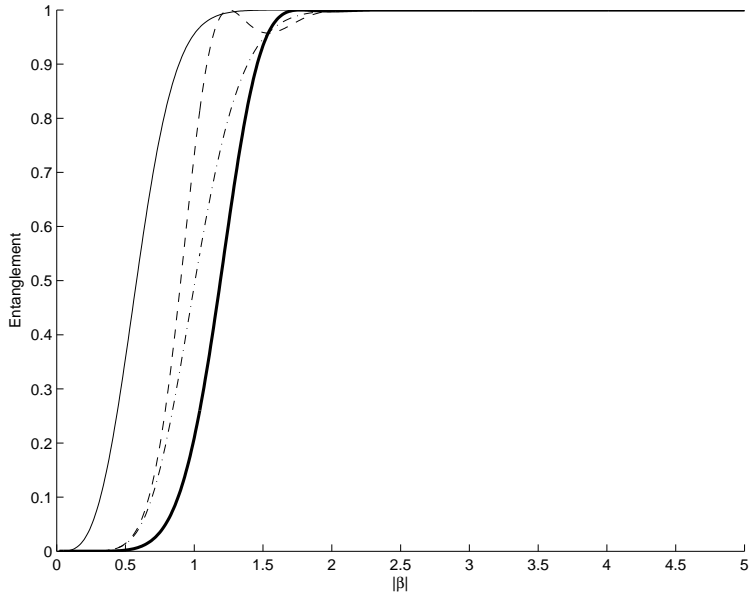}
\caption{Entanglement is shown as a function of $\vert\beta\vert$.  The curves correspond to $\phi=0.5\pi$.  The value of $\phi$ is chosen so that the dip in the entanglement in two-mode ECS is well pronounced.  The thin  continuous curve is for the two-mode coherent state, dashed curve for two-mode ECS, dashed-dotted curve for two-mode squeezed vacuum and the thick continuous line for two-mode OCS.}
\label{fig:Fig3}
\end{figure}
Among the states considered, the two-mode CS reaches close to one bit of entanglement with smaller value of $\vert\beta\vert$ compared to other states. The two-mode OCS attain the value of one bit of entanglement at a  much larger value of $\vert\beta\vert$. The two-mode ECS exhibits a sequence of maxima and minima  in entanglement whenever $\tanh(\vert\beta\vert^2)=\tan(\vert\beta\vert^2)$.  The first maximum occurs at $\vert\beta\vert\approx 1.25$  and the subsequent minimum at $\vert\beta\vert\approx 1.55$. Other extrema are not significantly large to be noticed in the figure.  Thus, the two-mode ECS provides an example of a state attaining one bit of entanglement for finite values of the amplitude.

The logarithmic states are described by specifying the values of $\gamma$, which is a complex number and $q$ whose value lies within the unit disc in the complex plane.  The condition for attaining the maximum in concurrence is expressed by the following  relation between the parameters $\gamma$ and $q$: 
\begin{equation}\label{qGammacondition}
\vert q\vert = \sqrt{\exp(\vert\gamma\vert^2)-1}.
\end{equation}
Since the magnitude of $q$ is limited to unity, value of $\vert\gamma\vert$ is restricted to $\approx 0.83255$. If $\gamma$ exceeds this limit, then the concurrence does not exhibit a peak.  Nevertheless, one bit of entanglement is attained as $\vert q\vert$ approaches unity.  These features are seen in the curves shown in Fig. 4.  The curve that corresponds to $\gamma = 0.1$ exhibits a peak in entanglement at $\vert q\vert = 0.1033$, as predicted by the relation in Eq. \ref{qGammacondition}.  There is no maximum in the curve corresponding to $\gamma =0.9$ as the corresponding value of $\vert q\vert$, satisfying Eq. \ref{qGammacondition}, exceeds unity.   If $\vert\gamma\vert\rightarrow\infty$, equivalently $\vert q\vert\rightarrow 0$,  $A_{LS}$ becomes unity making the concurrence and hence the entropy to vanish.  In this limit, the number state expansion of the LS involves only the two-mode vacuum state $\vert 0\rangle\vert 0\rangle$ and the state is separable. It is the case with the other generalized coherent states as well due to the fact that both $\beta$ and $(-1)^{1/k}\beta$ vanish when the amplitude is zero  and the superposition involves only the state $\vert m\rangle\vert m\rangle$.  
\begin{figure}[htp]
\centering
\includegraphics[width=10cm, height=8cm]{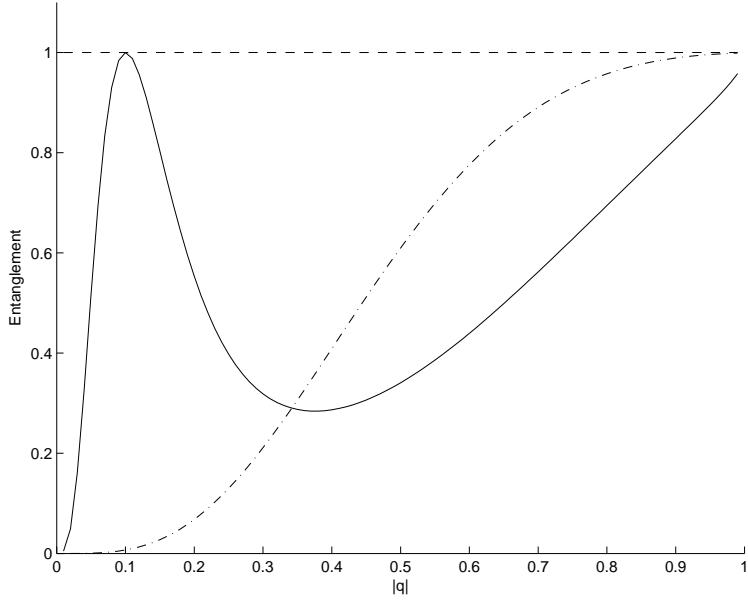}
\caption{Entanglement in two-mode logarithmic states is shown as a function of $\vert q\vert$.  The continuous curve corresponds to $\gamma=0.1$ and the dotted curve corresponds to $\gamma=0.9$. The later case does not exhibit a maximum as the condition in Eq. \ref{qGammacondition} is not satisfied. In both the cases $\phi=0.5\pi$.  The dashed curve, which remain constant at unity, corresponds to $\gamma=0.1$ and $\phi=\pi$.    }
\label{fig:Fig4}
\end{figure}

	Yet another type of superposition of two-mode generalized coherent states, linearly independent of the states defined in Eq.\ref{2modecs}, is
\begin{equation}\label{2modecs2}
N\left[\bcs\bncs+\exp(i\phi)\bncs\bcs\right].
\end{equation} 
The normalization constant  $N$ is $1/\sqrt{2[1+X^2\cos\phi]}$.  The relevant $a_j$ coefficients for these states are
\begin{eqnarray}
a_1 &=& NA^{-2}(1+\exp(i\phi)),\\
a_2 &=& -NB^{-2}(1+\exp(i\phi)),\\
a_3 &=& -NAB(1-\exp(i\phi)),\\
a_4 &=& NAB(1-\exp(i\phi)).
\end{eqnarray}
The concurrence, computed by substituting the coefficients $a_j$ given above into Eq.\ref{deter}, is the same as for the states specified in Eq. \ref{2modecs}.  Hence, all the properties of entanglement are identical for both the types of states.   

\section{Summary}
The amount of entanglement in the two-mode generalized coherent states is related to the parameter $A$.  All generalized  coherent states of equal $A$ possess equal entanglement.  When the parameter takes the value $\sqrt{2}$, the entanglement is exactly one bit irrespective of the relative phase. This general result is complementary to the result that the entanglement is one bit for all values of the amplitude when the relative phase is $\pi$.  In many cases, one bit of  entanglement is attained when the amplitude tends to infinity or the maximum allowed value.  However, there are examples, for instance, the logarithmic states, for which the condition is satisfied for amplitudes  which are lower than the allowed limiting value.

\end{document}